\documentclass{article}
\usepackage{spconf,amsmath,graphicx}
\usepackage{booktabs} 
\usepackage{multirow}
\usepackage{lscape}
\usepackage{todonotes}
\usepackage[acronyms]{glossaries}
\usepackage{amsmath}

\usepackage{mdframed}

\ninept

\makeglossaries
\newacronym{PSG}{PSG}{polysomnography}
\newacronym{OSA}{OSA}{obstructive sleep apnea}
\newacronym{CSA}{CSA}{central sleep apnea}
\newacronym{EEG}{EEG}{electroencephalogram}
\newacronym{EMG}{EMG}{electromyogram}
\newacronym{EOG}{EOG}{electrooculogram}
\newacronym{ECG}{ECG}{electrocardiogram}
\newacronym{PPG}{PPG}{photoplethysmography}
\newacronym{CASA}{CASA}{Computational Auditory Scene Analysis}
\newacronym{ITD}{ITD}{interaural time difference}
\newacronym{IID}{IID}{interaural intensity difference}
\newacronym{ISD}{ISD}{interaural spectral difference}
\newacronym{ASR}{ASR}{automatic speech recognition}
\newacronym{AIM}{AIM}{auditory image model}
\newacronym{SBE}{SBE}{snoring/breathing episode}
\newacronym{SNR}{SNR}{signal to noise ratio}
\newacronym{ROC}{ROC}{receiver operator characteristic}
\newacronym{PFLH}{PFLH}{Passion for Life Healthcare}
\newacronym{HMM}{HMM}{hidden Markov model}
\newacronym{GMM}{GMM}{Gaussian mixture model}
\newacronym{HTK}{HTK}{Hidden Markov Model Toolkit}
\newacronym{DTW}{DTW}{dynamic time warping}
\newacronym{SVM}{SVM}{support vector machine}
\newacronym{PCA}{PCA}{principal component analysis}
\newacronym{MFCC}{MFCC}{mel-frequency cepstral coefficient}
\newacronym{GCC}{GCC}{generalised cross-correlation}
\newacronym{NHS}{NHS}{National Health Service}
\newacronym{UARS}{UARS}{upper airway resistance syndrome}
\newacronym{DNN}{DNN}{deep neural network}
\newacronym{TP}{TP}{true positives}
\newacronym{TN}{TN}{true negatives}
\newacronym{FP}{FP}{false positives}
\newacronym{FN}{FN}{false negatives}
\newacronym{LM}{LM}{language model}
\newacronym{PM}{PM}{periodicity measure}
\newacronym{EER}{SER}{snore event error rate}
\newacronym{WER}{WER}{word error rate}
\newacronym{RM}{RM}{rate map}
\newacronym{ACF}{ACF}{autocorrelation function}
\newacronym{SDB}{SDB}{Sleep-disordered breathing}
\newacronym{CNN}{CNN}{convolutional neural network}

\title{DEEP LEARNING FEATURES FOR ROBUST DETECTION OF ACOUSTIC EVENTS IN SLEEP-DISORDERED BREATHING}
\name{Hector E. Romero, Ning Ma, Guy J. Brown,  Amy V. Beeston and Madina Hasan\thanks{H.E. Romero was funded by a PhD studentship from Passion for Life Healthcare (PFLH) and the Department of Computer Science, University of Sheffield. A.V. Beeston was funded by KTP award 9905 with PFLH.}}
\address{
  Department of Computer Science, University of Sheffield, Sheffield S1 4DP, UK\\
  {\small \tt \{heromeroramirez1, n.ma, g.j.brown, a.beeston, m.hasan\}@sheffield.ac.uk}
}

\begin{document}
\ninept
\maketitle
\begin{abstract}
Sleep-disordered breathing (SDB) is a serious and prevalent condition, and acoustic analysis via consumer devices (e.g. smartphones) offers a low-cost solution to screening for it. We present a novel approach for the acoustic identification of SDB sounds, such as snoring, using bottleneck features learned from a corpus of whole-night sound recordings. Two types of bottleneck features are described, obtained by applying a deep autoencoder to the output of an auditory model or a short-term autocorrelation analysis. We investigate two architectures for snore sound detection: a tandem system and a hybrid system. In both cases, a `language model' (LM) was incorporated to exploit information about the sequence of different SDB events. Our results show that the proposed bottleneck features give better performance than conventional mel-frequency cepstral coefficients, and that the tandem system outperforms the hybrid system given the limited amount of labelled training data available. The LM made a small improvement to the performance of both classifiers.
 
\end{abstract}

%


\section{Introduction}
\label{sec:intro}

\gls{SDB} is caused by the partial or complete collapse of the upper airway during sleep, whose forms include snoring, \gls{UARS}, and \gls{OSA}~\cite{dafna-paper}. Snoring is produced by the vibration of structures such as the soft palate, epiglottis, pharyngeal walls, and tongue, due to turbulent airflow caused by partial collapse of the upper airway~\cite{pevernagie-snoring-acoustics}. \gls{UARS} is indicated by abnormal respiratory effort originating from the limited airflow~\cite{guilleminault-UARS}. 
A complete collapse of the upper airway results in \gls{OSA}, characterised by an absence of airflow that interrupts sleep. \gls{OSA} has a strong relationship with cardiovascular \cite{lee-snoring-atherosclerosis}, metabolic, and neurocognitive diseases, affecting approximately 24-26\% of men and 9-28\% of women in Europe and the United States~\cite{sleep-apnea-review}, and its prevalence is increasing~\cite{bbc-sleep-testing}.

The current gold standard for diagnosing \gls{SDB} is \gls{PSG} ~\cite{dafna-paper, abeyratne-ANN, sleep-apnea-review, mendonca-home-osa}, which is expensive, time consuming, and uncomfortable for the patient. It involves sleeping for a complete night in a laboratory while physiological parameters are measured via at least 22 wired attachments to the body. A further problem is that data obtained from PSG in a hospital laboratory may not be representative of that which would be recorded in the home~\cite{abeyratne-ANN, sleep-apnea-review}. For this reason, alternatives to the diagnosis of \gls{SDB} have been explored including at-home \gls{PSG}~\cite{miller-home-SA-testing, mendonca-home-osa} and smartphone-based solutions \cite{apnea-app,snorelab}. In particular, smartphone apps offer the potential for a convenient, non-invasive, and low-cost method to diagnose \gls{SDB} through acoustic analysis~\cite{koo-smartphone-obstruction-site}.

The focus of this paper is the detection of \gls{SDB} sounds in whole-night audio recordings. Previous studies have typically used high-quality recordings made close to the patient's head while sleeping in clinic \cite{dafna-paper}; however a smartphone-based solution must work with higher levels of background noise, microphones that are designed for close-talking rather than ambient sound recording, and haphazard placement of the device. Recently, \glspl{DNN} have proven to be very effective in such sound classification tasks. However, \glspl{DNN} have not been fully exploited in \gls{SDB} sound classification because it is a low-resource task; there are few (if any) large corpora of labelled \gls{SDB} acoustic data. 

Previous studies have used a variety of acoustic features and classifier architectures. Nonaka et al. \cite{abeyratne-AIM} described a method to classify sleep audio recordings using a logistic regression classifier and features derived from the \gls{AIM}. In another study, Dafna et al. \cite{dafna-paper} used an AdaBoost classifier with 34 features, such as periodicity, total energy, duration, and higher-order spectral statistics. Two studies are notable for their use of deep learning. Firstly, Emoto et al. \cite{abeyratne-ANN} described a method for detecting low intensity snore and breathing events using a \gls{DNN}. Good accuracy was obtained on a subject-specific task (91.8\%) but performance was poorer on a subject-independent task (75.7\%), underlining the problem of generalising a \gls{DNN} when training data is limited. Secondly, Amiriparian et al. \cite{amiriparian2017} proposed a deep learning approach for snore sound classification. They overcame the low-resource problem by using a \gls{CNN} that was pre-trained on an image classification task to generate deep spectrum features, which were then classified by a \gls{SVM}.


The current paper introduces a number of innovations. First, we describe a corpus of \gls{SDB} acoustic data recorded via smartphones in domestic environments, and an annotation scheme for labelling that data. Second, in order to leverage a large volume of unlabelled data, we employ bottleneck features for SDB classification learned by a deep autoencoder. Noting that conventional audio features such as \glspl{MFCC} do not give a good representation of pitch, which might be important in SDB sound classification, we propose bottleneck features learned from the short-term \gls{ACF} in addition to those learned from an auditory representation of the sound spectrum. To further address the low-resource nature of this task, we investigate two snore detection architectures. The first is a \textit{tandem} approach in which a \gls{HMM}-\gls{GMM} system is used to model the acoustics of the bottleneck features. The second is a \textit{hybrid} approach in which a second \gls{DNN} is used to classify the bottleneck features. Finally, we investigate the utility of a `language model', obtained from our labelled data set, which captures information about the temporal sequence of different SDB events. 


The remainder of the paper is organised as follows. Section 2 describes the \gls{SDB} corpus and proposes an annotation scheme. Following this, the bottleneck features and classifier architectures are described in Section 3. An evaluation is presented in Section 4, using \glspl{MFCC} as a baseline for comparison. The final two sections of the paper discuss our results and make some concluding remarks. 



\section{Sleep breathing sound corpus}
\label{sec:corpus}
Given the lack of a suitable \gls{SDB} corpus, large-scale data collection was undertaken. Whole-night recordings were made of 31 male and 13 female participants in their own homes, using a custom app for iOS devices (e.g., iPhone, iPad or iPod Touch). The device was placed at head level within arm's reach, while the participant was normally sleeping. Audio recordings were single channel, sampled at 16 kHz with 16 bit depth. Data collection and storage protocols were subjected to the ethical review procedures of the University of Sheffield. 

\subsection{Data selection}

A subset of the data was selected for annotation. Since \gls{SDB} sounds such as snoring occur sporadically throughout a whole night, a simple \gls{GMM}-based classifier was implemented to identify signals of interest. Specifically, whole-night recordings were divided into 2-minute segments and those which contained at least 20\% of snore sound were identified, according to the automatic classification. From those, 25 segments were selected for each of six male participants, amounting to a total of 5 hours data. To balance the amount of data for other sound classes, such as background noise, a further 54 minutes of audio recordings were included which were made in a home environment when the participants were not sleeping. 

\subsection{Annotation}

Table \ref{table:annotation-scheme} shows the scheme used for annotating the \gls{SDB} sound recordings. In contrast to previous studies such as \cite{dafna-paper, abeyratne-AIM, abeyratne-ANN}, which only considered `snore' and `non-snore', here we adopt a more detailed annotation scheme. Specifically, six event types were defined for annotation: \textit{snore}, \textit{breath}, \textit{silence}, \textit{wheezing}, \textit{noisy in-breath}, and \textit{other}. This approach provides more flexibility, since similar event types can be merged if necessary. For the purposes of the snore detection experiments described in this paper, the \textit{snore}, \textit{wheezing}, and \textit{noisy in-breath} classes were merged into a single \textit{snore} class.

Each 2-minute signal was labelled by one of three annotators. To confirm the validity of this approach (i.e., to ensure that all annotators shared a common understanding of the annotation scheme), two audio recordings were first segmented and labelled by all three annotators, and compared at the frame level. A Cohen's kappa \cite{cohens-kappa} of 0.61$\pm$0.17 was obtained. Given the greater level of detail in our annotation scheme (i.e., six labels rather than two), this value suggests an acceptable level of agreement between annotators.

\begin{table}[t]
\vspace{-0.2cm}
\caption{Annotation scheme}
\vspace{.1cm}
\centering
\label{table:annotation-scheme}
\begin{tabular}{p{2cm} p{5.5cm}}
\toprule
\textbf{Acoustic event} & \textbf{Description} \\ \toprule
Snore & Pitched breathing sound. \\ \midrule
Wheezing & Whistling sound produced in the respiratory airways during breathing. \\ \midrule
Noisy in-breath & Like a snore, but without a strong pitch. \\ \midrule
Breath & Unpitched breathing sound, high frequency noise. \\ \midrule
Other & Speech, street noise, alarm clock, etc. \\ \midrule
Silence & Hiss, no other structure. \\ 
\bottomrule
\end{tabular}
\vspace{-.2cm}
\end{table}

\section{System Description}
\label{sec:features-and-classifiers}



Robust detection of acoustic events in \gls{SDB} is challenging because breathing sounds display great variability between individuals. 
Like speech, \gls{SDB} sounds vary according to the characteristics of an individual's vocal tract, and other factors can influence  breathing sounds during sleep (e.g., medical conditions such as asthma and emphysema \cite{thorax71}). Audio recordings made via smartphones in a home environment present further variabilities, due to the microphone used in different phones, placement of the device, and domestic noises.

For these reasons, various features and classifiers were studied with the objective of developing a robust \gls{SDB} classifier. The following two sections describe the proposed bottleneck features and classifier architectures.

\subsection{Bottleneck features}

Annotating the \gls{SDB} data is time-consuming: it takes on average 30 minutes to annotate a 2-minute section. The annotated corpus described in Section~\ref{sec:corpus} represents only a small portion of the data collected. To leverage the large amount of unlabelled data, we adopted an unsupervised approach by using a deep autoencoder. The autoencoder was trained on 20 hours of unlabelled data and extracts bottleneck features specialised for \gls{SDB} signals. These were derived from an auditory spectral representation and a pitch-based representation arising from the short-term \gls{ACF}.

\begin{figure}[thb]
\includegraphics[width=\linewidth]{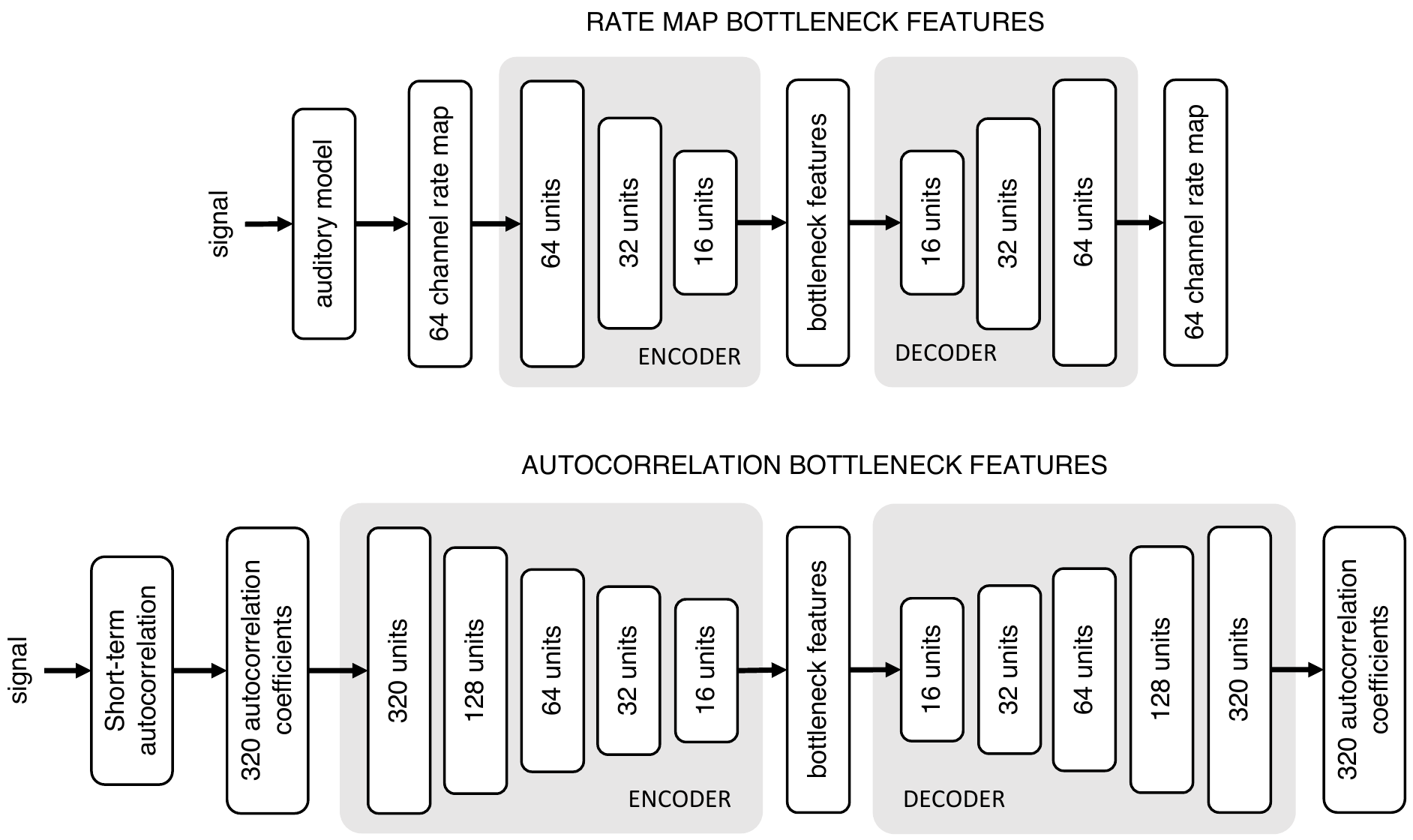}
\caption{Rate map and autocorrelation bottleneck features.}
\label{fig:acf-bottleneck-example}
\vspace{-.2cm}
\end{figure}

\subsubsection{Bottleneck features from auditory nerve firing rate maps}
\label{sec:bottleneck-rate-map}

Auditory-motivated representations have been successful in various sound understanding applications \cite{brown-CASA}. Here, we derived bottleneck features from a model of peripheral auditory processing based on a gammatone filterbank. 64 auditory filter channels were used, with centre frequencies spaced on the ERB-rate scale \cite{glasberg-erb} between 80 Hz and 7500 Hz. The envelope was then extracted from each channel, forming a so-called \gls{RM}. The envelope in each frequency channel can be interpreted as the instantaneous firing rate of an auditory nerve fibre \cite{brown-CASA}.


An autoencoder \gls{DNN} was implemented in TensorFlow \cite{tensorflow} to learn \gls{RM} bottleneck features. It consists of three fully connected layers that encode \glspl{RM}, followed by a further three fully connected layers that decode them (Fig.~\ref{fig:acf-bottleneck-example}, upper panel). During training, the aim is to reconstruct the input at the output layer, via a compressed intermediate representation. The input to the autoencoder is the 64-channel \gls{RM}, which is transformed to a compressed 16-channel representation by three layers of 64, 32 and 16 sigmoid units. After the \gls{RM} is encoded, it is decoded (i.e.,~reconstructed) back to its original 64-channel form by three layers of 16, 32 and 64 sigmoid units. The bottleneck features are obtained from the output of the 16-unit encoder layer; the second half of the network is discarded after training. Bottleneck features were appended with first-order and second-order difference (deltas and accelerations), resulting in a 48-element feature vector.


%
%

\subsubsection{Bottleneck features from the autocorrelation function}
\label{sec:bottleneck-acf}


Snoring is a \textit{pitched} acoustic event, since it is produced by the vibration of structures such as the soft palate, epiglottis and pharyngeal walls~\cite{pevernagie-snoring-acoustics}. The \gls{RM} bottleneck features, which can be seen as a compressed time-frequency representation, do not give a good representation of pitch, however. To provide pitch-related information to the system, we propose another kind of bottleneck features learned from the \gls{ACF}. The short-term \gls{ACF} is a popular means of pitch estimation~\cite{fundamental-frequency} and is defined as:
\begin{equation}
A(\tau) = \sum_{n=0}^{N-1} y(n)y(n-\tau)
\end{equation}
where $\tau$ is the time lag and $y$ is a windowed frame with $N$ samples. In this study $N=400$ for a 25\,ms window sampled at 16\,kHz. The first 320 lags were selected as input to the deep autoencoder, giving a lower pitch limit of 50 Hz.

The \gls{ACF} autoencoder is shown in the lower panel of Fig.~\ref{fig:acf-bottleneck-example}, which follows the same procedure as that of the \gls{RM} bottleneck features. A 320-lag \gls{ACF} is encoded to a 16-channel representation with five fully connected layers of 320, 128, 64, 32 and 16 sigmoid activation units, and decoded back to its original form with another five fully connected layers of 16, 32, 64, 128 and 320 sigmoid activation units. Again, the aim during training is to reconstruct the input at the output layer. The 16-channel encoded \gls{ACF} features were appended with first-order and second-order difference (deltas and accelerations), resulting in a 48-element feature vector as before.

\begin{figure}[t]
\centering
\includegraphics[width=\linewidth]{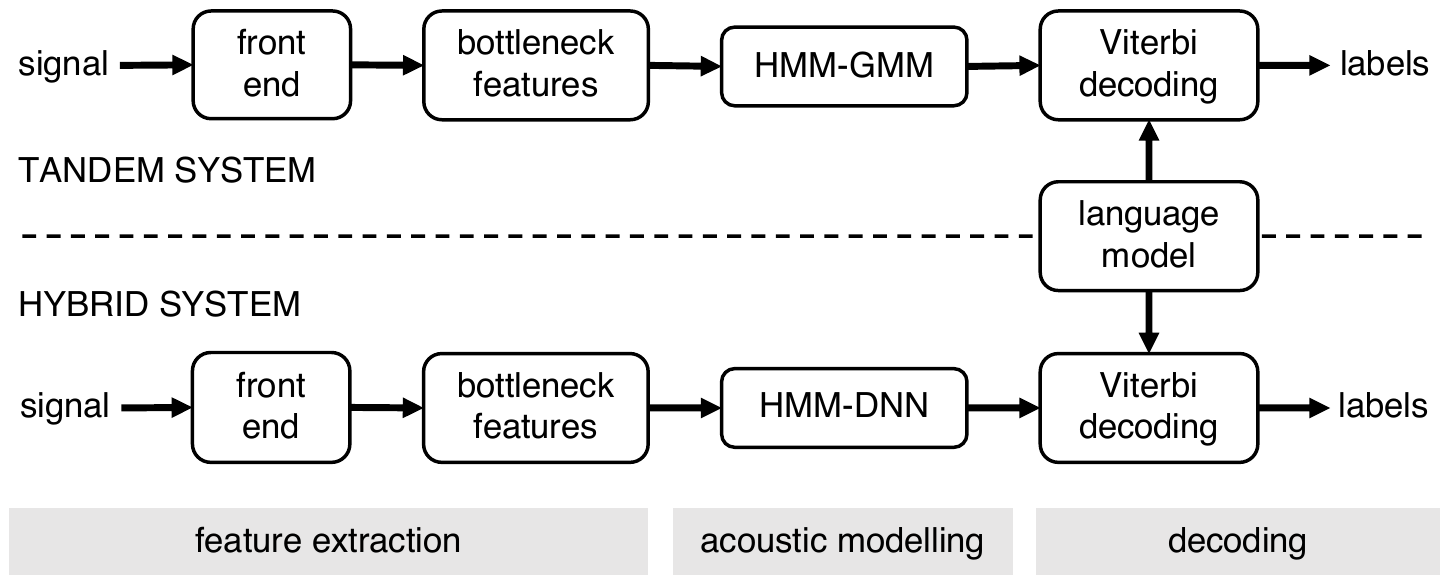}
\caption{System diagram showing both the tandem and the hybrid snore detection systems.}
\label{fig:system-diagram}
\vspace{-.2cm}
\end{figure}

\subsection{Snore detection architectures}



Given the limited amount of training data, two snore detection architectures were investigated using the bottleneck features, namely a \textit{tandem system} and a \textit{hybrid system}. A schematic overview of both systems is shown in Fig.~\ref{fig:system-diagram}.

\subsubsection{Tandem system}


In the tandem system, bottleneck features are extracted from the autoencoder \gls{DNN} but they are modelled by a conventional \gls{HMM} system with \glspl{GMM} modelling the state distributions~\cite{tandem-hybrid-systems}. The number of \gls{HMM} states  used by each class is summarised in Table~\ref{t:hmm_states}. Each state is represented by a \gls{GMM} with 7 Gaussian components using diagonal covariance matrices. The system is implemented using HTK~\cite{htk} and the parameters are selected heuristically according to the complexity of each class.

\begin{table}[thb]
\vspace{-.2cm}
\caption{Number of HMM states used in the tandem system}
\label{t:hmm_states}
\vspace{.1cm}
\centering
\begin{tabular}{c c c c c}
\toprule
\textbf{Class} & Snore & Breath & Other & Silence \\ 
\midrule
\textbf{States} & 7 & 5 & 3 & 3 \\
\bottomrule
\end{tabular}
\vspace{-.3cm}
\end{table}

\subsubsection{Hybrid system}



In the hybrid system, the HMM state distributions are directly represented by class posterior probabilities from a \gls{DNN}. The acoustic features are mapped by a \gls{DNN} with 3 hidden layers to the class labels. For the system that takes separate bottleneck features (RM bottleneck or ACF bottleneck) as input, each hidden layer consists of 96 fully connected hidden units with sigmoid activation functions. For the system that combines both bottleneck features, each hidden layer uses 192 hidden units. Finally, the output layer has four softmax units. This system is implemented using TensorFlow \cite{tensorflow}.


\subsection{Language model}

Breathing sounds, like language, usually follows certain patterns, e.g.,~a snore sound typically begins with an inspiration and is followed by a brief period of silence before an expiration. A \gls{LM}, such as is normally used in speech recognition, might therefore be leveraged for more accurate snore detection during the decoding process. 


HTK is used to compute event bigram statistics from the labelled training dataset. Given the limited amount of training data, the bigram model proves a suitable choice and is easy to incorporate. The \gls{LM} is employed by both the tandem and the hybrid systems during the decoding process.  In both systems, the Viterbi algorithm is employed to search for the most probable acoustic event sequence given the acoustic model and \gls{LM}.







\section{Evaluation}
\label{sec:evaluation}


\subsection{Experiment setup}

The annotated \gls{SDB} corpus described in Section~\ref{sec:corpus} was used for evaluation. The six-snorer data was divided into a training set which included all the annotated data from four snorers, and a test set consisting of data from the remaining two snorers. In this way, the systems we present here are `snorer-independent'.

For the tandem system, the \gls{HMM} parameters were estimated from the training data using the Baum-Welch algorithm. For the hybrid system, the \gls{DNN} was trained with a learning rate of 0.001, 60 epochs and a batch size of 256. During Viterbi decoding, an insertion penalty was used to balance the number of event deletion and insertion errors. Furthermore, a \gls{LM} scaling factor was also introduced for each system in order to adjust the impact of the \gls{LM}. Given the limited amount of annotated data, a randomly selected subset of the training data was used to empirically optimise the two parameters.

\subsection{Baseline systems}

\glspl{MFCC} have conventionally been used in \gls{ASR} and non-speech-related tasks such as rare sound event detection~\cite{event-detection-mfcc} and snore detection~\cite{duckitt-snore-detection}.
Both a tandem system and a hybrid system using \glspl{MFCC}, whose architectures are identical to the proposed systems, were introduced as the baseline systems. 12 \glspl{MFCC} excluding the C0 coefficient were extracted for each 25\,ms frame with a 10\,ms step. They were further appended with deltas and accelerations, forming 36-dimensional feature vectors.

\subsection{Evaluation metrics}

All systems were evaluated both at event level and frame level. At event level, the \gls{EER} was computed, which takes into account all the possible types of errors: insertions, deletions and substitutions, similar to the \gls{WER} commonly used in \gls{ASR}. 


At frame level, the snore $F$-measure was computed to evaluate the segmentation quality of the snore detection systems. This metric takes into account both the precision (proportion of frames classified as `snore' that are actually `snores') and the recall (proportion of actual `snore' frames that are classified as `snore'):

\begin{equation}
\label{eq:f-measure}
F\text{-measure} = \frac{2 \cdot \text{Precision} \cdot \text{Recall}}{\text{Precision} + \text{Recall}}
\end{equation}



%
%
%


\section{Results and Discussion}
\label{sec:results}

\begin{table}[t]
\vspace{-0.2cm}
\caption{Snore event error rates for various systems}
\vspace{0.1cm}
\centering
\label{table:snore-event-error-rate}
\begin{tabular}{@{}p{2.9cm} c c c c@{}}
\toprule
 & \multicolumn{2}{c}{\textbf{Tandem}} & \multicolumn{2}{c}{\textbf{Hybrid}} \\ 
 & \textbf{No LM} & \textbf{LM} & \textbf{No LM} & \textbf{LM} \\ \toprule
\textbf{MFCCs} & 19.94\% & 17.59\% & 9.40\% & 9.52\% \\ \midrule
\textbf{RM bottleneck} & 12.00\% & 12.13\% & 13.40\% & 13.40\% \\ \midrule
\textbf{ACF bottleneck} & 15.24\% & 14.48\% & 14.92\% & 14.92\% \\ \midrule
\textbf{RM+ACF bottleneck} & 10.86\% & \textbf{8.89\%} & 10.22\% & 9.90\% \\ 
\bottomrule
\end{tabular}
\vspace{-0.3cm}
\end{table}

The event-level \glspl{EER} and frame-level snore $F$-measures are shown in Tables \ref{table:snore-event-error-rate} and \ref{table:snore-f-measure}, respectively. Considering the \gls{EER}, the overall best performance was achieved by the tandem system using both RM and ACF bottleneck features with a \gls{LM} (\gls{EER}: 8.89\%), outperforming the baseline \gls{MFCC} system. The tandem systems using RM or ACF bottleneck features on their own produced significantly higher \glspl{EER}, which suggests the \gls{DNN} was able to exploit both the spectral and pitch information from the bottleneck features. A similar pattern of results can be seen for the hybrid systems. The use of the \gls{LM} improved the performance in most systems, especially in the tandem systems, which showed an average relative improvement of 9\%.



In most cases the tandem system outperformed the hybrid system. This is likely due to the amount of training data needed for the hybrid HMM-DNN system to perform well, while the HMM-GMM architecture adopted in the tandem system requires less data. This kind of behaviour has also been observed in other studies (e.g., \cite{gmm-hmm-dnn}). An exception is the baseline \gls{MFCC} system, which performed significantly better in the hybrid configuration (best \gls{EER}: 9.40\%) than in the tandem configuration  (best \gls{EER}: 17.59\%). This pattern is also observed in the frame-level results discussed below; the reason for this discrepancy is currently under investigation.



\begin{table}[t]
\vspace{-0.2cm}
\caption{Frame-based snore $F$-measures for various systems}
\vspace{0.1cm}
\centering
\label{table:snore-f-measure}
\begin{tabular}{@{}p{2.9cm} c c c c@{}}
\toprule
 & \multicolumn{2}{c}{\textbf{Tandem}} & \multicolumn{2}{c}{\textbf{Hybrid}} \\ 
 & \textbf{No LM} & \textbf{LM} & \textbf{No LM} & \textbf{LM} \\ \toprule
\textbf{MFCCs} & 90.78\% & 91.67\% & 93.60\% & 93.45\% \\ \midrule 
\textbf{RM bottleneck} & \textbf{95.29\%} & 95.23\% & 90.74\% & 90.74\% \\ \midrule
\textbf{ACF bottleneck} & 88.34\% & 88.47\% & 86.96\% & 86.96\% \\ \midrule
\textbf{RM+ACF bottleneck} & 94.43\% & 94.36\% & 94.73\% & 94.75\% \\
\bottomrule
\end{tabular}
\vspace{-0.3cm}
\end{table}

Looking at the frame-based snore $F$-measure as a way to assess the segmentation quality, the best results were achieved by the tandem configuration using only \gls{RM} bottleneck features ($F$-measure: 95.29\%), although the best result using both RM+ACF bottleneck features is very close ($F$-measure: 94.75\%). Comparing the $F$-measures and the \glspl{EER}, it is reasonable to hypothesise that a $F$-measure above 93\% demonstrates a reasonable quality in the segmentation, as a score below 93\% corresponds to a significant drop in the \gls{EER}. 


It should be noted that in the hybrid bottleneck systems, feature extraction and acoustic modelling are currently done in two separate \glspl{DNN}. This could potentially be a limitation. A better strategy might be to combine the two \glspl{DNN} into a single network, therefore allowing the bottleneck \gls{DNN} parameters to be adapted together with the acoustic \gls{DNN} model. 




\section{Conclusions}
\label{s:conc}

Robust snore detection in a home environment, from recordings made using a smartphone, is a challenging task. There are two major problems: (i) collection and annotation of a large amount of \gls{SDB} data; (ii) selection of appropriate acoustic features. We have described a solution to both problems by learning bottleneck features from a large corpus of unlabelled \gls{SDB} data, and then employing a tandem architecture that makes the most of a limited amount of labelled data. We obtained the best performance using bottleneck features that encode both spectral shape and pitch information; these were shown to outperform conventional \gls{MFCC} features when used in a tandem system. 

Detailed annotation of sleep breathing sounds is a surprisingly difficult task due to the subjectivity involved in the process, the inherent variability of breathing sounds, and the significant amount of time that it demands. For example, there is no commonly accepted acoustic definition of a snore \cite{pevernagie-snoring-acoustics}. In our experience, annotators require careful training if they are to adhere to agreed standards such as the one shown in Table~\ref{table:annotation-scheme}.

Although the contribution of a \gls{LM} was small for some systems that we investigated, it does provide useful information in the tandem system. Preliminary work suggests that the \gls{LM} helps to enforce realistic snore event durations. The \gls{LM} is likely to have a larger effect when identifying conditions that involve a temporal sequence of events, such as \gls{OSA}.

In the future we will focus on building systems to detect other forms of \gls{SDB}, such as \gls{UARS} and \gls{OSA}. Such systems will build on the snore detection approaches described here, for example by using a bag-of-audio-words approach~\cite{bag-audio-words} to combine snore event detection with other acoustic features over a whole night recording.


\vfill\pagebreak

\bibliographystyle{IEEEbib}
\bibliography{strings,refs}

\begin{thebibliography}{10}

\bibitem{dafna-paper}
E.~Dafna, A.~Tarasiuk, and Y.~Zigel,
\newblock ``Automatic detection of whole night snoring events using non-contact
  microphone,''
\newblock {\em PLoS ONE}, vol. 8, no. 12, December 2013.

\bibitem{pevernagie-snoring-acoustics}
D.~Pevernagie, R.~Aarts, and M.~De Meyer,
\newblock ``The acoustics of snoring,''
\newblock {\em Sleep Medicine Reviews}, vol. 14, pp. 131--144, 2010.

\bibitem{guilleminault-UARS}
T.~Masri and C.~Guilleminault,
\newblock ``Upper airway resistance syndrome,''
\newblock in {\em Encyclopedia of Sleep}. Academic Press, 2013.

\bibitem{lee-snoring-atherosclerosis}
G.~Lee, L.~Lee, C.~Wang, N.~Chen, T.~Fang, C.~Huang, W.~Cheng, and H.~Li,
\newblock ``The frequency and energy of snoring sounds are associated with
  common carotid artery intima-media thickness in obstructive sleep apnea
  patients,''
\newblock {\em Nature Scientific Reports}, 2016.

\bibitem{sleep-apnea-review}
M.~Shokoueinejad, C.~Fernandez, E.~Carroll, F.~Wang, J.~Levin, S.~Rusk,
  N.~Glattard, A.~Mulchrone, X.~Zhang, A.~Xie, M.~Teodorescu, J.~Dempsey, and
  J.~Webster,
\newblock ``Sleep apnea: a review of diagnostic sensors, algorithms, and
  therapies,''
\newblock {\em Physiological Measurement}, vol. 38, pp. 204--252, 2017.

\bibitem{bbc-sleep-testing}
BBC,
\newblock ``{Sleep disorder testing carried out by NHS doubles},''
  http://www.bbc.co.uk/news/uk-england-40122979, June 2017.

\bibitem{abeyratne-ANN}
T.~Emoto, U.~R. Abeyratne, K.~Kawano, T.~Okada, O.~Jinnouchi, and I.~Kawata,
\newblock ``Detection of sleep breathing sound based on artificial neural
  network analysis,''
\newblock {\em Biomedical Signal Processing and Control}, vol. 41, pp. 81--89,
  2018.

\bibitem{mendonca-home-osa}
F.~Mendon{\c c}a, S.~S. Mostafa, A.~G. Ravelo-Garcia, F.~Morgado-Dias, and
  T.~Penzel,
\newblock ``Devices for home detection of obstructive sleep apnea: A review,''
\newblock {\em Sleep Medicine Reviews}, 2018.

\bibitem{miller-home-SA-testing}
J.~Miller, P.~Schulz, B.~Pozehl, D.~Fiedler, A.~Fial, and A.~M. Berger,
\newblock ``Methodological strategies in using home sleep apnea testing in
  research and practice,''
\newblock {\em Sleep and Breathing}, November 2017.

\bibitem{apnea-app}
R.~Nandakumar, S.~Gollakota, and N.~Watson,
\newblock ``Contactless sleep apnea detection on smartphones,''
\newblock in {\em MobiSys 2015}, 2015, pp. 45--57.

\bibitem{snorelab}
SnoreLab,
\newblock ``{SnoreLab},'' https://snorelab.com, 2018.

\bibitem{koo-smartphone-obstruction-site}
S.~K. Koo, S.~B. Kwon, Y.~J. Kim, J.~S. Moon, Y.~J. Kim, and S.~H. Jung,
\newblock ``{Acoustic analysis of snoring sounds recorded with a smartphone
  according to obstruction site in OSAS patients},''
\newblock {\em European Archives of Oto-Rhino-Laryngology}, vol. 274, pp.
  1735--1740, 2017.

\bibitem{abeyratne-AIM}
R.~Nonaka, T.~Emoto, U.~R. Abeyratne, O.~Jinnouchi, I.~Kawata, H.~Ohnishi,
  M.~Akutagawa, S.~Konaka, and Y.~Kinouchi,
\newblock ``Automatic snore sound extraction from sleep sound recordings via
  auditory image modeling,''
\newblock {\em Biomedical Signal Processing and Control}, vol. 27, pp. 7--14,
  2016.

\bibitem{amiriparian2017}
S.~Amiriparian, M.~Gerczuk, S.~Ottl, N.~Cummins, M.~Freitag, S.~Pugachevskiy,
  A.~Baird, and B.~Schuller,
\newblock ``Snore sound classification using image-based deep spectrum
  features,''
\newblock in {\em Proceedings of Interspeech}, Stockholm, Sweden, 2017, pp.
  3512--3516.

\bibitem{cohens-kappa}
J.~Cohen,
\newblock ``A coefficient of agreement for nominal scales,''
\newblock {\em Educational and Psychological Measurement}, vol. 20, no. 1, pp.
  37--46, 1960.

\bibitem{thorax71}
P.~Forgacs, A.~R. Nathoo, and H.~D. Richardson,
\newblock ``Breath sounds,''
\newblock {\em Thorax}, vol. 26, no. 3, pp. 288--295, 1971.

\bibitem{brown-CASA}
G.~J. Brown and D.~L. Wang,
\newblock ``Fundamentals of computational auditory scene analysis,''
\newblock in {\em Computational Auditory Scene Analysis}, G.~J. Brown and D.~L.
  Wang, Eds., pp. 1--44. IEEE, 2006.

\bibitem{glasberg-erb}
B.~R. Glasberg and B.~C.~J. Moore,
\newblock ``Derivation of auditory filter shapes from notched-noise data,''
\newblock {\em Hearing Research}, vol. 47, pp. 103--138, 1990.

\bibitem{tensorflow}
M.~Abadi et~al.,
\newblock ``{TensorFlow: A system for large-scale machine learning},''
\newblock in {\em {Proceedings of the 12th USENIX Symposium on Operating
  Systems Design and Implementation}}. {USENIX}, 2016, pp. 265--283.

\bibitem{fundamental-frequency}
A.~de~Cheveigne and H.~Kawahara,
\newblock ``{YIN, a fundamental frequency estimator for speech and music},''
\newblock {\em J. Acoust. Soc. Am.}, vol. 11, no. 4, pp. 1917--1930, 2002.

\bibitem{tandem-hybrid-systems}
S.~P. Rath, K.~M. Knill, A.~Ragni, and M.~J.~F. Gales,
\newblock ``Combining tandem and hybrid systems for improved speech recognition
  and keyword spotting on low resource languages,''
\newblock in {\em {Interspeech Proceedings 2014}}. Interspeech, 2014.

\bibitem{htk}
S.~Young, G.~Evermann, M.~Gales, T.~Hain, D.~Kershaw, X.~Liu, G.~Moore,
  J.~Odell, D.~Ollason, D.~Povey, V.~Valtchev, and P.~Woodland,
\newblock {\em The HTK Book, HTK version 3.4 edition},
\newblock Cambridge University Engineering Department, 2006.

\bibitem{event-detection-mfcc}
J.~Gutierrez, R.~Fraile, A.~Camacho, T.~Durand, J.~Jarrin, and S.~Mendoza,
\newblock ``Synthetic sound event detection based on {MFCC},''
\newblock in {\em {DCASE proceedings}}. DCASE, September 2016.

\bibitem{duckitt-snore-detection}
W.~D. Duckitt, S.~K. Tuomi, and T.~R. Niesler,
\newblock ``Automatic detection, segmentation and assessment of snoring from
  ambient acoustic data,''
\newblock {\em Physiological Measurement}, vol. 27, pp. 1047--1056, 2016.

\bibitem{gmm-hmm-dnn}
J.~Schr{\"o}der, J.~Anem{\"u}ller, and S.~Goetze,
\newblock ``{Peformance comparison of GMM, HMM and DNN based approaches for
  acoustic event detection within task 3 of the DCASE 2016 challenge},''
\newblock in {\em {DCASE proceedings}}. DCASE, September 2016.

\bibitem{bag-audio-words}
M.~Schmitt, C.~Janott, V.~Pandit, K.~Qian, C.~Heiser, W.~Hemmert, and
  B.~Schuller,
\newblock ``A bag-of-audio-words approach for snore sounds' excitation
  localisation,''
\newblock in {\em {Speech Communication; 12. ITG Symposium}}, 2016, pp.
  230--234.

\end{thebibliography}

\end{document}